\documentstyle[preprint,aps]{revtex}
\begin{document}
\draft
\tighten

\title{An improved effective potential for electroweak phase
transitions}
\author{Janaki Balakrishnan and Ian G. Moss}
\address{Department of Physics, The University, Newcastle upon
Tyne, NE1 7RU, U.K.}
\maketitle
\begin{abstract}
It is shown that improved potentials and corrected mass terms
can be introduced by using a quadratic source term in the path
integral construction for the effective action. The advantage of
doing things this way is that we avoid ever having to deal with
complex propagators in the loop expansion. The resulting effective
action for electroweak phase transitions is similar to the usual
results.
\end{abstract}
\section{Introduction}

According to the standard models of cosmology and particle physics
the early stages of the universe were exceedingly hot and at least
some of the forces of nature were unified in a single force.
The symmetry governing this unified force would have been broken
when a Higgs field
developed a non-zero vacuum expectation value and the four
separate fundamental forces we observe today came into being as
a consequence of this.  As the standard electroweak theory has
been experimentally confirmed, it has become interesting to
study electroweak phase transitions at high temperatures (see
for instance \cite{sher,carrington}, as a preliminary
guide to understanding the physics of the early universe.

An important tool in discussing the dynamics of such phase
transitions in quantum field theory is the method of the
effective potential. This was first introduced by Goldstone,
Salam and Weinberg \cite{goldstone} and by Jona-Lasinio \cite{jona},
as a functional Legendre transform with an external source $J$
coupled linearly to the field $\phi$. It is possible to
calculate the effective action for the field of interest in
perturbation theory, by making an expansion in powers of
Planck's constant.

Although the effective potential formalism provides an efficient
way to get quantum corrections to the classical potential, it
requires some careful handling. The first problem that can arise
is that the perturbative expansion of the effective potential in
powers of $\hbar$ is not real when there is symmetry-breaking,
and for some values of the fields is even ill-defined. Secondly,
the exact effective potential according to some definitions is convex.
In a theory with spontaneous symmetry breaking at the classical level,
this means that the exact effective potential cannot have a double-well
shape and the local maximum of the system at the symmetric point at
the origin would never appear, even though the classical theory implies
the existence of a maximum at the origin.

An early suggestion for overcoming these problems was to introduce
a `corrected mass' into the propagator which is used in expanding
out the effective potential \cite{kirzhnits}. This is very similar
to a resummation of subsets of graphs  \cite{weinberg1,dine} to
obtain a corrected propagator. A difficulty with this approach
is that each individual diagram might itself be ill--defined,
and steps need to be taken to avoid this happening.

If, however, one introduced another source $K$ coupled to a term
which is quadratic in the fields in the definition of
the effective action, one would obtain an effective potential with a
proper loop expansion to each order which is not convex.
The idea of having
a quadratic source term for discussing the quantum field theory
of the field $\phi$ in the early universe was first put
forward by \cite{hawking} (see also \cite{amelino}).  This work was
based on an
important paper by Cornwall, Jackiw and Tomboulis \cite{cornwall}, who
defined an effective action for composite fields in flat space
as a double Legendre transform, with two sources $J(x)$ and $K(x,x')$
which were coupled respectively to $\phi(x)$ and $\phi(x)\phi(x')$. An
application of these ideas was made recently in \cite{moss,moss1},
where a new effective action was defined for scalar fields at finite
temperature and for an $O(N)$ gauge theory at zero
temperature.

The derivation of the effective action in \cite{moss} differs
from the conventional effective action by having both sources, $J$ and
$K$ present initially, but the source $J$ is made to vanish
in the limit of the spatial volume going to infinity.
The key point of this procedure is the condition
that the expectation values are dominated by field configurations
that lie along the direction of $J$ before the limit is taken.
The expectation values are calculated with both sources present
initially, but when the limit $J\rightarrow0$  and volume
$\Omega\to\infty$ is taken, the magnitude of the
fields retain only their dependence on the source $K$ and lead to real
masses.  The beauty of this procedure is that, whereas on the one hand
the conventional effective action is regained by setting $K=0$ when
there is no symmetry breaking, on the
other hand keeping $K$ finite and setting $J\rightarrow0$ at the
end gives a well-defined effective action for broken symmetric
theories.
Moreover, this effective action is not
necessarily convex, which means that it can have the double well
shape of its classical theory.

For a system which is confined to a finite volume, the
configuration space is divided into subsets with equal
probability measure which are related by symmetry
transformations. In the infinite volume limit, if one of the
configuration space subsets develops a higher probability for
its measure than the others, then the system exhibits symmetry
breaking. This is the standard procedure for setting up a pure
state in a theory having only the linear source term. The problem
with this from our point of view is that the volume of the
system may always be finite, because of cosmological considerations,
and that it does not address the issues concerned with dynamically
evolving fields.

In the alternative procedure used here\cite{moss,moss1},
the pure broken symmetric state is prepared in the same way as
before, with $J = 0$ in the infinite volume limit, and
also by setting $K=0$.
In addition, however, the system can also be prepared
in a state located at the local symmetric maximum near the
origin of field space by setting $J=0$ as before but by keeping
$K$ non-zero and finite. This is why the exact effective potential
is not necessarily convex in this approach. All this is valid at zero
as well as at non--zero temperature.

Interestingly, the improved effective potential (IEP)
described in \cite{moss} gives the same (positive)
mass squared of the scalar field at equilibrium at high
temperatures as that obtained by Kirzhnits and Linde \cite{kirzhnits}.

In this work, we shall develop the method described in \cite{moss} and
show in Section 3 that in this new approach, one can reproduce
two important results of the background field method, namely
that the effective field equations are given by
\begin{equation}
{\delta\Gamma[\phi]\over\delta \phi}=0
\end{equation}
and that the effective action defined
in \cite{moss} is invariant under BRS transformations $\hat s$ of
the fields, in the sense that
\begin{equation}
\langle \hat s\Gamma\rangle=0.
\end{equation}
Finally in Sections 4 and 5, we shall use this
definition to calculate the effective potential for the Higgs
field in the standard $SU(2) \times U(1)$
electroweak theory. At zero temperatures the resulting potential
is real, without any resummation of graphs, and gives the usual
mass relations for the Higgs, $W$ and $Z$ bosons in terms of
the Higgs expectation value. At non--zero temperatures our results
compare favourably with others \cite{dine,arnold}.

\section{IMPROVED EFFECTIVE ACTIONS}

First we shall very briefly review the method of the improved
effective potential for a scalar field described in \cite{moss}.
We shall use Riemannian signature (++++) for the metric throughout and
work in units in which $c=1$.

Consider a real scalar field $\Phi$, which we will allow to have more
than one internal component. The composite field partition function
$Z[J,K]$ is defined by a path integral \cite{cornwall},
\begin{eqnarray}
Z[J,K] &=& e^{-W[J,K]/\hbar}\nonumber\\
&=&\int d\mu[\Phi] \exp\left\{-I_{JK}[\Phi]\right\},
\end{eqnarray}
where
\begin{equation}
I_{JK}[\Phi] =I[\Phi] +
\int d\mu(x) \Phi(x)^TJ(x) +
\case1/2\int d\mu(x) d\mu(x')\,\Phi(x)^T K(x,x')
\phi(x')\label{z},
\end{equation}
and $I[\Phi]$ is the action. If $\Phi$ has $N$ components, then $K$ is
an $N\times N$ matrix.

The expectation value of the field $\phi$ and the connected
propagator $G$ are defined in the presence of the external sources by
\begin{eqnarray}
\phi(x)&=&{\delta W[J,K]\over\delta J(x)}  \nonumber\\
\hbar G(x,x')&=&
2{\delta W[J,K]\over\delta K(x,x')}-
{\delta W[J,K]\over\delta J(x)}{\delta W[J,K]\over\delta J(x')}.
\end{eqnarray}
It is also possible to write $G(x,x')$ as,
\begin{equation}
\hbar G(x,x') = \langle\Phi(x)\Phi^T(x')\rangle -
\langle\Phi(x)\rangle\langle\Phi^T(x')\rangle \label{k}
\end{equation}
The composite effective action $\Gamma[\phi,G]$ is then defined by
a double Legendre transform,
\begin{eqnarray}
\Gamma[\phi,G] = W[J,K] - \int d\mu(x) \phi^T(x)J(x) -
\case1/2\int d\mu(x) d\mu(x') \phi(x)^T K(x,x') \phi(x')\nonumber\\
- \case{\hbar}/{2}\int d\mu(x) d\mu(x') G(x,x')K(x',x)
\end{eqnarray}
The effective field equation in the presence of the two sources
$J$ and $K$ follow from functional differentiation of the effective
action,
\begin{eqnarray}
{\delta \Gamma[\phi,G]\over\delta \phi(x)} &=& - J^T(x) -
\int d^4y \phi^T(y) K(y,x) \label{jj}\\
{\delta \Gamma[\phi,G]\over\delta G(x,x')} &=&
-\case1/2\hbar K(x,x')\label{kk}
\end{eqnarray}
Physical on-shell solutions are obtained by setting the sources
$J$ and $K$ to zero in (\ref{jj}) and (\ref{kk}). When $K$ is
zero, eqn.(\ref{kk}) can be solved for $G$ in terms of $\bar\phi$, and
this value of $G$ can be substituted back in (\ref{jj}) to get the
conventional effective action,  $\Gamma[\phi] =
\Gamma[\phi,G(\phi)]$.

However, $G$ need not necessarily be obtained only in this
manner. There are other possible solutions for $G$, ones with
non-zero $K$, which can be substituted into $\Gamma[\phi]$
to get a different form of $\Gamma[\phi,G(\phi)]$.
The conventional definition of the effective action for $\phi$
considers only the special case $K=0$. The presence of $K$
allows the effective action to give information
not only about the ground state but also about the symmetric
solution about the origin.

In order to proceed further it will be helpful to examine what
the field configurations are that dominate the partition function.
We will take the classical symmetry breaking potential
\begin{equation}
V(\Phi) = -\case1/2\mu^2 {|\Phi|}^2 +
\case1/4\lambda{|\Phi|}^4
\end{equation}
In the first place consider what happens with a constant linear
source,
\begin{equation}
J(x)=j,
\end{equation}
and diagonal quadratic source,
\begin{equation}
K(x,y) = k\,{\rm I}\,\delta(x-y).
\end{equation}

The path integral is dominated by saddle points, obtained from the
condition
\begin{equation}
{\delta I_{JK}\over\delta\Phi} = 0\label{sp},
\end{equation}
which gives
\begin{equation}
V'(\phi_s) + j + k\phi_s = 0               .
\end{equation}
This is solved for small $j$ by making the ansatz
\begin{equation}
\phi = \sum_n a_n |j|^n.
\end{equation}
The first few terms in the series are
\begin{equation}
\phi_{s\pm} = \pm{(\mu^2-k)^{1/2}\over\sqrt
\lambda} \hat \jmath - \case1/2 {(\mu^2 -k)}^{-1} j + O(j^2).
\end{equation}
There are therefore two solutions, one in the direction of $j$ and
one in the opposite direction. If $j=0$, then instead of two solutions
there is a continuum of saddle points with a fixed value of the
modulus $|\phi|$.

We will proceed now without assuming that $J(x)$ is constant.
The total partition function $Z[J,K]$ should be constructed by summing
the contributions from all of the
saddle points separately. For example,
\begin{equation}
Z \sim e^{-W_+/\hbar} + e^{-W_-/\hbar}\label{dec}
\end{equation}
where
\begin{equation}
e^{-W_+[J,K]/\hbar}=e^{-I_{JK}[\phi_+]/\hbar}
\int d\mu[\tilde\phi]\exp\left(-I^{(1)}/\hbar\right).
\end{equation}
The action $I^{(1)}$ consists of the terms from the expansion of
$I_{JK}[\Phi]$, where $\Phi=\phi_++\tilde\phi$, that are linear or
higher order in $\tilde\phi$. This can be written,
\begin{equation}
I^{(1)}=\int d\mu(x)\tilde\phi^T(x) J^{int}(x)
+\int d\mu(x,x')\tilde\phi^T(x)(\Delta(x,x')
+ K(x,x'))\tilde\phi(x') +I^{(3)},
\end{equation}
with
\begin{equation}
J^{int}(x)={\delta I_{JK}\over\delta\Phi(x)}\hbox{,\quad}
\Delta(x,x')={\delta^2 I\over\delta\Phi^T(x)\delta\Phi(x')}\label{jint}.
\end{equation}

The partition function is obtained by summing the two series
obtained from each of the two saddle points. In practice, it
proves even better to expand about background fields $\phi_\pm$
shifted from the saddle points by $O(\hbar)$ quantum corrections.
Let
\begin{equation}
\langle\Phi_\pm(x)\rangle={\delta W_\pm\over\delta J(x)},
\end{equation}
then we solve
\begin{equation}
\phi_\pm=\langle\Phi_\pm(x)\rangle\label{tad}
\end{equation}
to get $\phi_\pm$.
Because of (\ref{sp}), $\phi=\phi_{s\pm}$ to leading order in $\hbar$.

By definition, the expectation value $\phi$ of the
field $\Phi$ is
\begin{equation}
\phi = -{\delta\over\delta J} \ln Z
\end{equation}
This can be rewritten using (\ref{dec}) as
\begin{equation}
\phi = {e^{-W_+/\hbar}\over Z}\phi_+  +
{e^{-W_-/\hbar}\over Z}\phi_-
\end{equation}
or alternatively
\begin{equation}
\phi=  \case1/2(\phi_+ + \phi_-) - \case1/2 (\phi_+
-\phi_-) {\rm tanh} ({W_+ - W_-\over2\hbar}).
\end{equation}
For constant sources, the expectation value
$\phi$ of the field $\Phi$ can be shown to be
\begin{equation}
\phi = -\case1/2{(\mu^2-k)}^{-1} j -
\frac{{(\mu^2-k)}^{\frac{1}{2}}}{\sqrt \lambda}
{\rm tanh}(\frac{W_+-W_-}{2\hbar})\hat \jmath  + O(j^2)
\end{equation}

As for the connected propagator $G(x,x')$, from eqn.(\ref{k})
\begin{equation}
\hbar G(x,x') = \langle \Phi(x)\Phi^T(x')\rangle -
\langle\Phi(x)\rangle\langle\Phi^T(x')\rangle.
\end{equation}
The asymptotic expansions about $\phi_\pm$ would then be,
\begin{equation}
\hbar G(x,x') \sim Z^{-1}e^{-W_+/\hbar}
{\langle\Phi(x)\Phi^T(x')\rangle}_+ + Z^{-1}e^{-W_-/\hbar}
{\langle\Phi(x)\Phi^T(x')\rangle}_--\phi(x)\phi^T(x').
\end{equation}
Some rearrangements of this result give,
\begin{equation}
\hbar G(x,x')
= \case1/2\hbar(G_+ +G_-) -
\case1/2\hbar(G_+-G_-){\rm tanh} \left(W_+-W_-\over2\hbar\right) +
\rho_c(x,x')
\end{equation}
where
\begin{equation}
 G_\pm = {\langle{\tilde \phi}(x)\tilde \phi(x')^T\rangle}_\pm
\end{equation}
and
\begin{equation}
\rho_c(x,x') = \case1/4 {\rm sech}^2\left(\frac{W_+-W_-}{2\hbar}
\right) {(\phi_+(x) -\phi_-(x))} (\phi_+(x') -\phi_-(x'))^T.
\end{equation}

In order to obtain an effective potential $\Gamma[\phi]$ it is necessary
to eliminate the propagator $G$. If we set the source $K=0$ the result
should be the usual effective action. There are two possible cases to
consider depending on whether we keep both of the series expansions
about $\phi_\pm$ or just one. With both series the result is a convex
potential \cite{moss}. Taking the infinite volume limit with
nonzero $J$ leads to the dominance of only one of the two series,
but in this case the propagator becomes
\begin{equation}
G\sim G_+\sim-\nabla^2+V''(\phi)
\end{equation}
to order $\hbar$. The mass term in this propagator is imaginary for
small values of $\phi$.

The alternative way of eliminating $G$ is to take
\begin{equation}
K(x,x')=k(x)\delta(x,x')
\end{equation}
and $J\to0$. However, when $J=0$ there is a continuum of saddle points
in the path integral. This would cause problems with the asymptotic
expansions and therefore we arrange that only one saddle point
should dominate the result. This can be done by taking the limit
$\Omega \phi.J\to-\infty$ as the volume $\Omega\to\infty$. This limit
can be imposed simultaneously with the limit $J\to0$.

The effective action can now be evaluated with only one saddle point
contribution corresponding to $\phi=\phi_+$. The composite field
effective action is defined by
\begin{eqnarray}
\Gamma[\phi,G] = W[J,K] - \int d\mu(x) \phi^T(x)J(x) -
\case1/2\int d\mu(x) d\mu(x') \phi^T(x) K(x,x') \phi(x')\nonumber\\
- \case1/2\hbar\int d\mu(x) d\mu(x') G(x,x')K(x,x')
\end{eqnarray}

The asymptotic series for $W$ can be expressed in terms of Feynman
diagrams with scalar field propagator $G_0=(\Delta+K)^{-1}$. This has
the form
\begin{equation}
W[K]=I[\phi_+]-\case1/2\hbar\ln\det\,G_0+W^{(2)},
\end{equation}
where $W^{(2)}$ contains all of the connected diagrams in the series
that have two or more loops. The full connected propagator $G(x,x')$
is given by
\begin{equation}
\hbar G(x,x')=\langle\tilde\phi(x)\tilde\phi^T(x')\rangle_+.
\end{equation}

Substituting $W$ into to definition of the action gives
\begin{equation}
\Gamma[\phi]\sim I[\phi] -\case1/2\hbar{\rm tr}(G_0K) -
\case1/2\hbar\ln\det\,G_0+W^{(2)}.
\end{equation}
Note that $K[\phi]$ is given by the condition (\ref{tad}), which
corresponds to the vanishing of the tadpole diagrams.

Following Cornwall et.al. \cite{cornwall}, it is possible
to rewrite the expression on the right in terms of the full propagator,
\begin{equation}
\Gamma[\phi]\sim I[\phi]-\case1/2\hbar{\rm tr}\left(1-G\Delta\right) -
\case1/2\hbar\ln\det\,G+\Gamma^{(2)}.
\end{equation}
where $\Gamma^{(2)}$ consists of two--particle irreducible diagrams.
The form of $G[\phi]$ is determined by equating the tadpole diagram
series to zero, with vertices representing the shifted theory
including $J^{int}$  (eqn.\ref{jint}).

\section{GAUGE THEORIES}

In this section we will set up the improved effective action for
Higgs fields in gauge theories. Our aim is to prove two results.
First of all we will show that the effective field equations are
given by variation of the effective action. Then we will show that
the effective action has the usual behaviour under BRS symmetry.
This is important because of the way that BRS symmetry underlies
proofs of renormalisability. Higgs and gauge fields are denoted by
$\Phi$ and $A$.

We will fix the gauge symmetry of the Lagrangian density by
adding gauge fixing and ghost terms,
\begin{equation}
{\cal L}_{tot}={\cal L}_\phi+{\cal L}_A+{\cal L}_{gf}+{\cal L}_{gh}.
\end{equation}
The approach that we take is essentially a background field expansion
where the gauge fixing Lagrangian density is likely to have an
explicit background field dependence. We denote these background
fields by $\hat\phi$ and $\hat A$, and we have dependencies
\begin{eqnarray}
{\cal L}_{gf}[\phi,A,\hat\phi,\hat A]\\
{\cal L}_{gh}[\phi,A,c,\bar c,\hat\phi,\hat A]
\end{eqnarray}
where $c$ and $\bar c$ are the ghost fields.

Fixing the gauge leaves a residual BRS symmetry $s$, whose action is
given by
\begin{eqnarray}
sA_\mu&=&D_\mu c,\nonumber\\
s\Phi&=&ic\Phi,\nonumber\\
sc&=&-\case1/2[c,c],\nonumber\\
s\bar c&=&\xi^{-1}{\cal F},\nonumber\\
s\hat A_\mu&=&[c,A_\mu],\nonumber\\
s\hat\phi&=&ic\hat\phi.
\end{eqnarray}
The gauge covariant derivative with connection $A$ is denoted by
$D$ and ${\cal F}$ is a gauge fixing functional.

External sources will be introduced for each of the quantum fields, but
for simplicity only the Higgs field will have a quadratic source.
We define,
\begin{equation}
I_{JK}=\int d\mu(x)\left\{
{\cal L}_{tot}+\Phi^T J+J^\mu A_\mu+\bar\theta c+\bar c\theta\right\}
+\case1/2\int d\mu(x,x')\Phi(x)^TK(x,x')\Phi(x').
\end{equation}

The generating functional is now
\begin{equation}
e^{-W/\hbar}=\int d\mu[\Phi,A,c,\bar c]e^{-I_{JK}}
\end{equation}
with connected diagrams generated by
$W[J,K,\theta,\bar\theta,\hat\phi,\hat A]$. The path integral
is evaluated perturbatively, expanding the Higgs field about
a background $\phi_+$. Later we will set $\phi_+$ to equal
the other background field $\hat\phi$. We will also arrange
that the other saddle points never contribute to the path
integral in the same way as in the previous section.

Construction of the effective action for composite fields proceeds
as before,
\begin{eqnarray}
\Gamma[\phi,A,c,\bar c,g;\hat\phi,\hat A]=W
-&&\int\left\{J\phi+J^\mu A_\mu+\bar\theta c+\bar c\theta\right\}
\nonumber\\
-&&\case1/2\int\phi^TK\phi-\case1/2{\rm tr}(KG).
\end{eqnarray}
Variation of this action gives back the sources,
\begin{eqnarray}
{\delta\Gamma\over\delta\phi(x)}&=&
-J^T(x)-\int d\mu(x')\phi^T(x')K(x',x)\label{eom}\\
{\delta\Gamma\over\delta G(x,x')}&=&-\case1/2K(x,x')\\
{\delta\Gamma\over\delta A_\mu(x)}&=&-J^\mu(x).
\end{eqnarray}

At this stage we would like to set $\hat\phi=\phi$ and eliminate
$G(x,x')$ to obtain an effective action that is a function of
the field expectation values and nothing else. The obstacle to
doing this is that there is an explicit dependence on $\hat\phi$
from the gauge fixing terms in the action. This would not be so
if the action where truly independent of the gauge fixing term,
but this is possible only in some approaches \cite{dewitt}.
A simple way to fix up this
problem is to subtract off the offending term,
\begin{equation}
\bar\Gamma=\Gamma-
\int\left\{\langle{\cal L}_{gf}\rangle+
\langle{\cal L}_{gh}\rangle\right\}.
\end{equation}
Now the variation of $\Gamma$ with $\hat\phi$ is cancelled by the
variation of the other two terms, and we may set
\begin{equation}
\Gamma[\phi,A,c,\bar c]=
\bar\Gamma[\phi,A,c,\bar c,G[\phi];\phi,A]
\end{equation}
after solving for $G[\phi]$ by using vanishing tadpole diagrams.
Finally, it follows from (\ref{eom}) that the effective field
equations becomes $\delta\Gamma=0$.

In low--order perturbation theory we can simplify the subtraction
of ${\cal L}_{gf}$ by making use of the $\xi$ dependence,
\begin{equation}
\int\langle{\cal L}_{gf}\rangle=\xi{d W\over d\xi}.
\end{equation}
In particular, at order $\hbar$ the $\xi$ dependence is linear and
therefore the subtraction is not necessary in the Landau gauge $\xi=0$.

The other issue of importance is gauge invariance. In a
background field approach it should be possible to retain
covariance under gauge transformations of the
background fields explicitly at each stage of the calculation.
Therefore the non--trivial issue is the residual BRS invariance left
after the gauge invariance of the quantum fields has been
fixed.

It should be possible to change the variables of integration
in the path integral to their images under a BRS transformation
without changing the generating functions. This implies that
the path integral of the BRS variation vanishes, or
\begin{equation}
\langle s\,I_{JK}\rangle=0\label{w}.
\end{equation}
The BRS symmetry of the action imples that
\begin{equation}
s\,I_{JK}=\int\left\{J s\Phi+J^\mu sA_\mu
+\bar\theta sc+s\bar c\theta\right\}+
\case1/2{\rm tr}\left(K s(\Phi\Phi^T)\right).
\end{equation}
The sources can be eliminated by expressing them as functional
derivatives of the (composite) effective action. We also define a new
transformation $\hat s$ that acts on expectation values and
produces quantum fields. For field expectation values it acts in the
same way as $s$, but the action of $\hat s$ on the propagator is
\begin{equation}
\hat s\,G=s\,\Phi\,\Phi^T+\Phi\,s\Phi^T-2\phi\,\hat s\Phi.
\end{equation}

Using the transformation $\hat s$ it is possible to write (\ref{w})
in the form
\begin{equation}
\langle \hat s\,\Gamma \rangle=0.
\end{equation}
This equation is an expression of the Ward--Slavnov--Taylor identities
for the composite effective action.

\section{IMPROVED EFFECTIVE ACTION FOR THE
STANDARD ELECTROWEAK THEORY }

We now compute the effective action at high temperature in
$SU(2) \times U(1)$ theory using the procedure developed in the
earlier sections, starting with the bosonic sector.

The classical Lagrangian density for the Glashow-Salam-Weinberg
model \cite{glashow,salam,weinberg} is given by
\begin{equation}
{\cal L}_{GSW} = {\cal L}_s + {\cal L}_g + {\cal L}_{gf} +
{\cal L}_f + {\cal L}_y +{\cal L}_{gh}.
\end{equation}
Leaving aside the fermion terms ${\cal L}_f$ and Yukawa
couplings ${\cal L}_y$
for the moment, we have,
\begin{eqnarray}
{\cal L}_s &=& {(D_\mu \Phi)}^\dagger (D^\mu \Phi) - \mu^2
\Phi^\dagger \Phi + \case1/2\lambda{(\Phi^\dagger \Phi)}^2,
\nonumber\\
{\cal L}_g &=& -\case1/2 \delta^{ab}F_{a \mu\nu}F^{\mu\nu}_b.
\end{eqnarray}
The conventions for the field strengths and covariant derivatives
which will be used are as follows. The field strength tensor
$F_{\mu\nu}=F_{\mu\nu a}T^a$, with
\begin{equation}
F_{\mu\nu} = \partial_\mu A_\nu  - \partial_\nu A_\mu
+ g[A_\mu, A_\nu]
\end{equation}
The covariant derivative $D_\mu$ is defined as
\begin{equation}
D_\mu = \nabla_\mu - ig A_{\mu ~a}T^a
\end{equation}
where
\begin{eqnarray}
T^a &=& \sigma^a  \hbox{\rm\quad for\quad}a=1,2,3\nonumber\\
T^a &=& {g'\over g}{\rm I} \hbox{\rm\quad for\quad}a=4.
\end{eqnarray}
The first three generators are Pauli matrices and the fourth
is proportional to the unit matrix $I$. Group indices can
be raised with the metric $2\delta^{ab}$ or lowered with
$\case1/2\delta_{ab}$.

A t'Hooft gauge fixing term will be used,
\begin{equation}
{\cal L}_{gf}=-\case1/2\xi^{-1}{\cal F}_a{\cal F}^b
\end{equation}
where
\begin{equation}
{\cal F}=\nabla\cdot A+i\xi g\hat\phi^\dagger T_a\Phi\,T^a.
\end{equation}
The corresponding ghost action is then
\begin{equation}
{\cal L}_c=\bar c^a\left(\nabla^2\delta_a^{\ b}
-\xi g^2\hat\phi^\dagger T_aT^b\Phi\right)c_b.
\end{equation}

Calculation of the effective potential begins by expanding the
action about a background in powers of the perturbed fields. The
only non--vanishing background field which we will consider will
be the Higgs field $\phi$,
\begin{equation}
\phi={1\over\sqrt{2}}\pmatrix{0\cr v\cr}.
\end{equation}
Perturbed fields will be denoted by a tilde. The total action is then
\begin{equation}
I[\phi+\tilde\phi,\tilde A,\tilde c,\bar{\tilde c}]=
\int d\mu(x)\left\{{\cal L}_2+{\cal L}_3+{\cal L}_4\right\}
\end{equation}
where ${\cal L}_i$ denotes the part of $\cal L$ that has the $i$th
power of the perturbed fields.

The quadratic terms determine fluctuation operators $\Delta$, given
explicitly by
\begin{eqnarray}
\Delta_A&=&-\delta_\mu^{\ \nu}\nabla^2+
(1-\xi^{-1})\nabla_\mu\nabla^\nu+X_A\delta_\mu^{\ \nu},\\
\Delta_\phi&=&-\nabla^2+X_\phi\\
\Delta_c&=&-\nabla^2+\xi X_A
\end{eqnarray}
where
\begin{eqnarray}
X_A&=&2g^2\phi^\dagger T_aT^b\phi\\
X_\phi&=&\xi g^2T_a\phi\phi^\dagger T^a-
\mu^2+\lambda\phi^\dagger\phi+2\lambda\phi\phi^\dagger.
\end{eqnarray}

The mass matrices can all be decomposed with respect to
the broken symmetry subgroup. A general way to do this is
described in appendix A. This gives,
\begin{eqnarray}
X_A&=&\case1/2 g^2v^2 P_{\parallel a}^{\ \ b}\\
X_\phi&=&(-\mu^2+\lambda v^2+\case1/2\xi g^2 v^2){\rm I}
+\lambda v^2 P_\parallel\label{xphi}
\end{eqnarray}

The effective action given in the previous section was
\begin{equation}
\Gamma=I-\case1/2\hbar{\rm tr}(1-G_\phi\Delta_\phi) +\Gamma^{(1)}
+\Gamma^{(2)}
\end{equation}
with
\begin{equation}
\Gamma^{(1)}=-\case1/2\hbar\ln\det G_\phi
-\case1/2\hbar\ln\det G_A+\hbar\ln\det G_c.
\end{equation}
The propagators are given by Feynman diagram expansions with
internal lines representing $(\Delta_\phi+K)^{-1}$ etc.

For constant background fields we can find the effective
potential from $V(\phi)=\Omega^{-1}\Gamma[\phi]$. We will use
Landau gauge $\xi\to0$, in which the ghosts are independent of the
background field. We also define corrected masses in terms of the
full propagators by,
\begin{eqnarray}
(G_\phi)^{-1}&=&-\nabla^2+X_\phi(T)+O(\hbar)\nonumber\\
(G_A)^{-1}&=&-\nabla^2\delta_\mu^{\ \nu}+X_A(T)+O(\hbar).
\end{eqnarray}

The one--loop contribution to the effective potential is then,
\begin{eqnarray}
V^{(1)}&&=-\case1/2\hbar\left(G_\phi^{-1}-\Delta_\phi\right)G_\phi(x,x)
+\case1/2\hbar\Omega^{-1}\ln\det(-\nabla^2+X_\phi(T))\nonumber\\
&&+\case1/2\hbar\Omega^{-1}{\ln\det}_T(-\nabla^2+X_A(T)).
\end{eqnarray}
The vector field determinant is over transverse modes.

The value of $k$ which appears in $G_0$ is found by equating
the tadpole diagram series to zero,
\begin{equation}
k=\mu^2-\case1/2\lambda v^2+O(\hbar^{-1}\lambda T^2)+O(\hbar)
\end{equation}
The leading order terms in the propagator are given by
\begin{equation}
X_\phi(T)=X_\phi-k\,{\rm I}+O(\hbar^{-1}\lambda T^2),
\end{equation}
with $X_\phi$ given in (\ref{xphi}). As in ref. \cite{moss},
the one--loop contribution to the propagator cancels with
the one--loop contribution to $k$. Thus,
\begin{eqnarray}
X_\phi(T)&=&m_\phi^2 P_\parallel+O(\hbar)\nonumber\\
X_A(T)&=&m_W^2 P_{\parallel a}^{\ b}+O(\hbar^{-1} g^2T^2)
\end{eqnarray}
where
\begin{equation}
m_\phi^2=\lambda v^2,\hbox{\quad}
m_W^2=\case1/2 g^2 v^2,\hbox{\quad}
m_Z^2=\case1/2 (g^2+g'^2)^{1/2} v^2
\end{equation}
The gauge loop corrections to the vector boson masses can be found
in the literature, e.g. \cite{carrington}.

The effective potential now becomes
\begin{eqnarray}
V^{(1)}&&\sim-\case1/2\hbar\left(\mu^2-\case1/2 m_\phi^2
\right)G_\phi(x,x)
+\case1/2\hbar\Omega^{-1}\ln\det(-\nabla^2+m_\phi^2)\nonumber\\
&&+\case3/2\hbar\Omega^{-1}
\left[2{\ln\det}_T(-\nabla^2+m_W^2)+
{\ln\det}_T(-\nabla^2+m_Z^2)\right].
\end{eqnarray}

This result is valid for any temperature. At zero temperature the
main feature of this result is the simple form of the Higgs boson
mass terms, which are generally different from those in the standard
one--loop result. The results agree, however, when the Higgs field
takes its vacuum expectation value.

In the high temperature limit, when $T>>m_\phi$ and $T>>m_W$, we
have the expansions,
\cite{dolan,weinberg1}
\begin{equation}
G(x,x)\sim\hbar^{-2}\left(\case1/{12}T^2-\case1/{2\pi}\hbar m T
-\case1/{8\pi^2}\hbar^2m^2\ln(\mu_R/T)\right)
\end{equation}
and
\begin{equation}
\ln\det G\sim\Omega\hbar^{-4}\left(\case{\pi^2}/{45}T^4
-\case1/{12}\hbar^2m^2T^2+\case1/{6\pi}\hbar^3m^3T
+\case1/{16\pi^2}\hbar^4m^4\ln(\mu_R/T)\right).
\end{equation}
The quantity $\mu_R$ is the renormalisation scale.

In this limit the effective potential becomes
\begin{eqnarray}
V^{(1)}\sim&&\case1/{12\pi}\mu^2 m_\phi T
+\case1/{16}\hbar^{-1}m_\phi^2 T^2
-\case1/{8\pi}m_\phi^3T+\hbar\case1/{16\pi^2}
(\mu^2-m_\phi^2)m_\phi^2\ln(\mu_R/T)\nonumber\\
&&+\case1/8\hbar^{-1}(2m_W^2+m_Z^2)T^2
-\case1/{4\pi}(2m_W^3+m_Z^3)T\nonumber\\
&&-\hbar\case3/{32\pi^2}(2m_W^4+m_Z^4)\ln(\mu_R/T).
\end{eqnarray}
The terms involving vector boson masses have the standard form
\cite{carrington,dine,arnold}. We have not included the ring corrections
to the vector boson masses, but these would be identical to those
in the references. The other terms have some resemblance
to results that use a corrected mass for the Higgs field,
the main difference with some of these is the term linear in
$v$. Although the authors of ref. \cite{dine} argue strongly
against such a term, their arguments are only valid for their
definition of the effective action.

\section{FERMION CONTRIBUTIONS}

As in the paper by Carrington \cite{carrington}, we work in the
approximation in which the only non-vanishing Yukawa coupling is
the top quark coupling constant, $h^t$.  The fermion part of the
Lagrangian is given by
\begin{equation}
{\cal L}_f = \bar\psi_R \gamma^\mu (\nabla_\mu -i\case1/2g'Y
B_\mu ) \psi_R + \bar\psi_L \gamma^\mu (\nabla_\mu
-ig\case1/2\sigma^aA_{\mu ~a} - i\case1/2g'Y B_\mu )\psi_L
\end{equation}
where the left-handed fermions have been denoted by $\psi_L$ :
\begin{equation}
\psi_L = \case1/2(1-\gamma_5)\pmatrix{\nu_\alpha\cr
e_\alpha\cr} ,\hbox{\quad} \case1/2(1-\gamma_5)
\pmatrix{q_{u \alpha}\cr q_{l \alpha}\cr}_{3~colours}
\end{equation}
and the right-handed fermions by $\psi_R$  :
\begin{equation}
\psi_R = \case1/2(1+\gamma_5) e_\alpha,
\hbox{\quad} \case1/2
(1+\gamma_5) q_{u \alpha},\hbox{\quad}
\case1/2(1+\gamma_5)q_{l \alpha}
\end{equation}
and $B=A_4$.

The subscript $\alpha$ denotes the three generations of quarks
$(u,d)$, $(c,s)$, and $(t,b)$ or for the leptons
$(\nu_e,e)$, $(\nu_\mu,\mu)$, $(\nu_\tau,\tau)$.
We have denoted the hypercharge by $Y$ and we follow the
notation in which the charge $Q$ of the field is defined by
$Q = T^3 + \case1/2Y$ , where $T^3$ is the third component of
isospin. In this notation, $Y=-1$ for the left handed leptons,
$Y=\case1/3$ for the left-handed quarks, $Y=-2$ for the right-handed
electrons, $Y=\case4/3$ for the right handed quarks $q_{u \alpha}$
and $Y=-\case2/3$ for $q_{l\alpha}$.

The Yukawa term ${\cal L}_y$ is given by
\begin{equation}
{\cal L}_y = - h^t \bar q_L \phi_c t_R
\end{equation}
where $\phi_c$ has been defined as
\begin{equation}
\phi_c = i \tau^2 \phi^*
\end{equation}

In our notation, the Dirac matrices $\gamma^a$ are all hermitean:
\begin{equation}
\gamma^0 = \pmatrix{0&{\rm I}\cr {\rm I}&0\cr},\ \
\gamma^i =\pmatrix{0&i\tau^i\cr-i\tau^i&0\cr},\ \
 \gamma^5 = \pmatrix{{\rm I}&0\cr 0&-{\rm I}\cr}
\end{equation}
with $i=1\dots3$.

In terms of the components of the Higgs field, the Yukawa term
can be rewritten as
\begin{eqnarray}
{\cal L}_y &&= -{h^t\over\sqrt 2}\bar t (\tilde\phi_3+v) t -
{h^t\over2\sqrt 2} \{~\bar b(1+\gamma_5)(-\tilde\phi_1
+i\tilde\phi_2)t\nonumber\\
&& - \bar t(1-\gamma_5)(\tilde\phi_1+i\tilde\phi_2)b ~\} + i
{h^t\over\sqrt 2}\bar t \gamma_5 \tilde\phi_4 t
\end{eqnarray}
where we have performed the shift in the Higgs field .

The fermionic contribution to the one-loop effective action is given
formally by
\begin{equation}
\Gamma^{(1)}_f = - \ln ~{\rm det}~D_f =  -\ln ~{\rm det}~\left[
 \gamma^\mu  \nabla_\mu - m_t \right]
\end{equation}
where we have denoted the top quark mass by
\begin{equation}
m_t = \frac{h^t
v}{\sqrt 2}.
\end{equation}
Since this determinant is ill-defined, we rewrite it in
terms of the determinant of $D^\dagger_f D_f$ which is
well defined, as
\begin{eqnarray}
\Gamma^{(1)}_f &=& -\case1/2 \ln {\rm det} ~D^\dagger_f D_f
\nonumber\\
&=& -\case1/2\ln {\rm det} ~( -\nabla^2 + m_t^2 )
\end{eqnarray}

The high temperature effective potential can be calculated as
in \cite{dolan}, or by using the result for fermions that
\begin{equation}
\left[{\ln\det}\Delta\right]_{antiperiod\ \beta} =
\left[2\,{\ln\det}\Delta\right]_{period\ 2\beta}
-\left[{\ln\det}\Delta\right]_{period\ \beta}
\end{equation}
where $\beta = 1/T$, and we work in units in which
Boltzmann's constant, $k=1$.
This gives,
\begin{equation}
\ln\det G\sim\Omega\hbar^{-4}\left(-\case7/8\case{\pi^2}/{45}T^4
+\case1/{24}\hbar^{-2}m^2T^2+\case1/{16\pi^2}
\hbar^{-4}m^4\ln(\mu_R/T)\right)
\end{equation}
and therefore
\begin{equation}
V^{(1)}_{fermions} = \hbar\left(\case1/4 {\hbar}^{-2}T^2m_t^2 +
\case3/8m_t^4 \ln (\mu_R/T) \right)
\end{equation}
where we have neglected all the terms of order $T^{-1}$ at high
temperatures.\\
We have also performed the trace over the colour and
the spin indices. It should be noted that the term linear in $T$
is absent for the case of fermions.

\section{CONCLUSIONS}

We have shown that improved potentials and corrected mass terms
can be introduced by using a quadratic source term in the path
integral construction. The advantage of doing things this way is
that we avoid ever having to deal with complex propagators in the
loop expansion. We are also able to include the possibility of
multiple saddle points in the path integaral, and in a
cosmological context where the volume might be finite this should
not be overlooked.

The resulting effective potential at zero temperatures is different
from the usual results. However, at finite temperatures, where it
has become standard to use corrected mass terms, the results are
mostly in agreement with the literature. Differences in the Higgs
contribution may be due to differences in truncation of the series.
If the linear terms that we have found are really present, then
the phase transition from the symmetric to the broken symmetry
phase is affected.

The phase transition takes place by bubble nucleation. These bubbles
satisfy the effective field equations generated by $\delta\Gamma=0$.
We have been careful to ensure that these equations are a property
of the effective action construction, even to the extent of
correcting for the gauge fixing terms.

\appendix
\section*{DECOMPOSITIONS OF MASS MATRICES}

The generators of the Lie algebra of the gauge group are represented
by matrices $T^a$. These define a metric $g^{ab}={\rm tr}(T^aT^b)$.
For product groups this metric is not unique because the relative
scaling of some generators is not fixed. Therefore another metric
$\tilde g^{ab}$ will be used to raise or lower indices.

The background Higgs fields $\phi$ can be used to define a projection
$P_\parallel$ by $\phi^\dagger \phi P_\parallel=\phi\phi^\dagger$,
and an orthogonal projection $P_\perp=1-P_\parallel$. The matrices
$P_\perp T^a P_\perp$ generate another Lie algebra. We define
\begin{eqnarray}
g_\perp^{ab}&=&{\rm tr}(P_\perp T^a P_\perp T^a P_\perp),\\
g_1^{ab}&=&{\rm tr}(P_\parallel T^a P_\parallel T^a P_\parallel),\\
g_\parallel^{ab}&=&2{\rm tr}(P_\parallel T^a P_\perp T^a P_\parallel).
\end{eqnarray}

These are not independent because
$g^{ab}=g_\perp^{ab}+g_1^{ab}+g_\parallel^{ab}$. We also define
operators that act on the Lie algebra
\begin{equation}
P_{\perp a}^{\ \ b}=\tilde g_{ac}(g_\perp^{cb}-g_1^{cb})\hbox{,\quad}
P_{\parallel a}^{\ \ b}=\tilde g_{ac}(g_\parallel^{cb}+2g_1^{cb}).
\end{equation}
The significance of these operators is that the vector mass matrix
in the Higgs model is
$2 g^2\phi^\dagger T_aT^b\phi=
g^2\phi^\dagger\phi P_{\parallel a}^{\ \ b}$.

For the group $SU(2)\times U(1)$ with doublet Higgs fields the
generators can be chosen to be $T^a=\sigma^a$ for $a=1\dots3$ and
$T^4={\rm I}\,t$, where $t=g'/g$. We take $\tilde g^{ab}=2\,\delta^{ab}$.
For $\phi=(0,v)^T/\sqrt{2}$, the corresponding projections are
\begin{equation}
P_{\perp a}^{\ \ b}=\pmatrix{0&&&&\cr &0&&&\cr &&0&t\cr &&t&0\cr}
\hbox{,~~~}
P_{\parallel a}^{\ \ b}
=\pmatrix{1&&&&\cr &1&&&\cr &&1&-t\cr &&-t&t^2\cr}.
\end{equation}

\end{document}